\documentstyle[natbib,epsfig,longtable]{aipproc}

\newcommand{\ee}{{\rm e^+e^-}}
\newcommand{\mm}{ \mu^+\mu^-}
\newcommand{\lum}{\cal{L}}

\newcommand{\Lunits}{\,{\rm cm^{-2}.s^{-1}}}
\newcommand{\ECoM}{{\rm E_{CoM}}}

\begin{document}

\setcounter{secnumdepth}{2}    

\pagestyle{plain}
\footskip 1.5 cm

\title{Parameter Sets for 10 TeV and 100 TeV Muon Colliders,
and their Study at the HEMC'99 Workshop
\thanks{
To appear in Proc. HEMC'99 Workshop  -- Studies on Colliders and
Collider Physics at the Highest Energies: Muon Colliders at 10 TeV to
100 TeV; Montauk, NY, September 27-October 1, 1999,
web page http://pubweb.bnl.gov/people/bking/heshop.
This work was performed under the auspices of
the U.S. Department of Energy under contract no. DE-AC02-98CH10886.}
}

\author{Bruce J. King}
\address{Brookhaven National Laboratory\\
email: bking@bnl.gov\\
web page: http://pubweb.bnl.gov/people/bking}
\maketitle

\begin{abstract}
  A focal point for the HEMC'99 workshop was the evaluation of
straw-man parameter sets for the acceleration and collider rings of
muon colliders at center of mass energies of 10 TeV and 100 TeV.
These self-consistent parameter sets are presented and discussed.
The methods and assumptions used in their generation are described
and motivations are given for the specific choices of parameter values.
The assessment of the parameter sets during the workshop is then reviewed
and the implications for the feasibility of many-TeV muon colliders are
evaluated. Finally, a preview is given of plans for iterating on the
parameter sets and, more generally, for future feasibility studies on
many-TeV muon colliders.
\end{abstract}

\section{Introduction}
\label{sec:intro}


   Self-consistent example parameter sets for the acceleration and collider
ring parameters of many-TeV muon colliders were an important focal point for
the discussions at the HEMC'99 Workshop -- ``Studies on Colliders and
Collider Physics at the Highest Energies: Muon Colliders at 10 TeV to
100 TeV'', held at Montauk, NY from September 27-October 1, 1999. They
served as straw-man examples to be criticized, fleshed-out and improved
upon by the accelerator experts attending the workshop, and the
physics-related parameters helped the experimental and theoretical
physicists at the workshop in their evaluations and comments on the
physics potential of such colliders.

  Three acceleration and collider parameter sets were used at HEMC'99:
one at a center-of-mass energy of 10 TeV (set A) and two at 100 TeV
(sets B and C). The collider ring and accelerator parameters are
presented in tables~\ref{colliderpara} and~\ref{accel}, respectively.
For comparison, table~\ref{colliderpara} also includes the parameter
ranges for the lower energy muon colliders that have been studied
by the Muon Collider Collaboration (MCC).
This paper describes the methods used to generate the parameter sets,
details the motivations and assumptions for the specific choices of
parameters and summarizes the evaluations, conclusions and suggestions
for the parameter sets that were given by the workshop participants.

  In more detail, the collider ring parameters are presented first, in
section~\ref{sec:collider}, since they were considered the more critical
of the two for assessing the feasibility of many-TeV muon colliders. They
also determine the initial assumptions used for the acceleration parameters,
which are then discussed in section~\ref{sec:accel}. The level of
understanding advanced substantially during the workshop, and
section~\ref{sec:assess}
goes over the issues and viewpoints raised during the workshop as well
as referencing the more detailed studies that are included elsewhere
in these proceedings and discussing their impact
on our assessment of the parameter sets. Finally, the Outlook and
Conclusions section, section~\ref{sec:concl}, summarizes the results
discussed in the preceding section in the more general context of what
they imply for the feasibility of many-TeV muon colliders. This
concluding section also discusses the outlook for iterations and
refinements on the parameter sets and, more generally, previews some
plans for further studies on many-TeV muon colliders.

\section{Straw-man Muon Collider Ring Parameter Sets at 10 TeV and 100 TeV}
\label{sec:collider}


%
%
%

\begin{table}[htb!]
\centering
\squeezetable
\renewcommand\tabcolsep{5pt}
\caption{
Self-consistent collider ring parameter sets for many-TeV muon colliders.
The parameters are as evaluated in the HEMC'99 workshop with the exception
of the neutrino radiation parameters, which have been updated to incorporate
the improved estimates from reference~\cite{hemc99nurad}.}
\begin{tabular}{|r|cccc|}
\hline
\multicolumn{1}{|c|}{ {\bf parameter set} }
                            &     & A  & B  & C \\
\multicolumn{1}{|c|}{ {\bf center of mass energy, ${\rm E_{CoM}}$} }
                            & 0.1 to 3 TeV & 10 TeV  & 100 TeV  &  100 TeV \\
\multicolumn{1}{|c|}{ {\bf additional description} }
                            & MCC status report & evol. extrap. & evol. extrap. & ultracold beam\\
\hline \hline
\multicolumn{1}{|l|}{\bf collider physics parameters:} & & & & \\
luminosity, ${\cal L}$ [${\rm 10^{35}\: cm^{-2}.s^{-1}}$]
                                        & $8 \times 10^{-5}$$\rightarrow$0.5
                                        & 10 & 10 & 1000 \\
$\int {\cal L}$dt [${\rm fb^{-1}/year}$]
                                        & 0.08$\rightarrow$540 & 10 000
                                        & 10 000 & $1.0 \times 10^6$ \\
No. of $\mu\mu \rightarrow {\rm ee}$ events/det/year
                                        & 650$\rightarrow$10 000 & 8700 & 87 & 8700 \\
No. of 100 GeV SM Higgs/year            & 4000$\rightarrow$600 000
                                        & $1.4 \times 10^7$
                                        & $2.1 \times 10^7$
                                        & $2.1 \times 10^9$ \\
CoM energy spread, ${\rm \sigma_E/E}$ [$10^{-3}$]
                                        & 0.02$\rightarrow$1.1 & 0.42 & 0.080 & 0.071 \\
\hline
\multicolumn{1}{|l|}{\bf collider ring parameters:}   & & & & \\
circumference, C [km]                   & 0.35$\rightarrow$6.0 & 15 & 100 & 100 \\
ave. bending B field [T]                & 3.0$\rightarrow$5.2 & 7.0 & 10.5 & 10.5 \\
\hline
\multicolumn{1}{|l|}{\bf beam parameters:}            & & & & \\
($\mu^-$ or) $\mu^+$/bunch, ${\rm N_0[10^{12}}]$
                                        & 2.0$\rightarrow$4.0 & 3.0 & 0.80 & 0.19 \\
($\mu^-$ or) $\mu^+$ bunch rep. rate, ${\rm f_b}$ [Hz]
                                        & 15$\rightarrow$30 & 27 & 7.9 & 65 \\
6-dim. norm. emit., $\epsilon_{6N}
               [10^{-12}{\rm m}^3$]     & 170$\rightarrow$170 & 85 & 10 & $1.0 \times 10^{-3}$\\
                    $\epsilon_{6N}
               [10^{-4}{\rm m}^3.{\rm MeV/c}^3$]
                                 & 2.0$\rightarrow$2.0 & 1.0 & 0.12 & $1.2 \times 10^{-5}$\\
P.S. density, ${\rm N_0}/\epsilon_{6N}
               [10^{22}{\rm m}^{-3}$]   & 1.2$\rightarrow$2.4 & 3.5 & 8.0 & 19 000 \\
x,y emit. (unnorm.)
              [${\rm \pi.\mu m.mrad}$]  & 3.5$\rightarrow$620 & 0.81 & 0.018 & $4.4\times10^{-4}$ \\
x,y normalized emit.
              [${\rm \pi.mm.mrad}$]     & 50$\rightarrow$290  & 38 & 8.7 & 0.21 \\
long. emittance [${\rm 10^{-3}eV.s}$]           & $0.81\rightarrow24$ & 21 & 47 & 8.1 \\ 
fract. mom. spread, $\delta$ [$10^{-3}$]
                                        & 0.030$\rightarrow$1.6 & 0.60 & 0.113 & 0.100 \\
relativistic $\gamma$ factor, ${\rm E_\mu/m_\mu}$
                                        & 473$\rightarrow$14 200  & 47 300
                                        & 473 000 & 473 000 \\
time to beam dump,
          ${\rm t_D} [\gamma \tau_\mu]$ & no dump & no dump & 1.0 & 1.0 \\
effective turns/bunch                  & 450$\rightarrow$780 & 1040 & 1350 & 1350 \\
ave. current [mA]                      & 17$\rightarrow$30 & 55 & 4.0 & 7.8 \\
beam power [MW]                        & 1.0$\rightarrow$29 & 131 & 100 & 198 \\
synch. rad. critical E [MeV]                   & $5 \times 10^{-7}$$\rightarrow$$8 \times 10^{-4}$
                                        & 0.012 & 1.75 & 1.75 \\
synch. rad. E loss/turn [GeV]                  & $7 \times 10^{-9}$$\rightarrow$$3 \times 10^{-4}$
                                        & 0.017 & 25 & 25 \\
synch. rad. power [MW]                    & $1\times10^{-7}$$\rightarrow$0.010
                                        & 0.91 & 99 & 195 \\
beam + synch. power [MW]                & 1.0$\rightarrow$29 & 130 & 200 & 390 \\
power density into magnet liner [kW/m]  & 1.0$\rightarrow$1.7 & 4.3 & 1.2 & 2.4 \\
\hline
\multicolumn{1}{|l|}{\bf interaction point parameters:}      & & & & \\
spot size, $\sigma_{x,y}$
                          $[\mu {\rm m}]$   & 3.3$\rightarrow$290 & 1.3 & 0.21 & 0.015 \\
bunch length, $\sigma_z$ [mm]          & 3.0$\rightarrow$140 & 2.2 & 2.5 & 0.49 \\
$\beta^*_{x,y}$ [mm]      & 3.0$\rightarrow$140 & 2.1 & 2.5 & 0.49 \\
ang. divergence, $\sigma_\theta$
                             [mrad]    & 1.1$\rightarrow$2.1 & 0.63 &  0.086 & 0.030 \\
ip compensation factor: ${\rm N_0/N_{0,eff.}}$  & 1 & 1 & 1 & 10 \\ 
beam-beam tune disruption, $\Delta \nu$
                                        & 0.015$\rightarrow$0.051 & 0.085 & 0.100 & 0.100 \\
pinch enhancement factor, ${\rm H_B}$  & 1.00$\rightarrow$1.01
                                        & 1.08 & 1.11 & 1.11 \\
beamstrahlung frac. E loss/collision  & negligible
                                       & $6.8 \times 10^{-8}$
                                       & $1.5 \times 10^{-6}$
                                       & $9.0 \times 10^{-7}$
                                           \\
\hline
\multicolumn{1}{|l|}{\bf final focus lattice parameters:} & & & & \\
max. poletip field of quads., ${\rm B_{5\sigma}}$ [T]
                                        & 6$\rightarrow$12 & 15 & 20 & 20 \\
max. full aper. of quad., ${\rm A_{\pm5\sigma}}$[cm]
                                        & 14$\rightarrow$24 & 22 & 19 & 6.6 \\
quad. gradient,    $2{\rm B_{5\sigma} / A_{\pm5\sigma}}$[T/m]
                                        & 50$\rightarrow$90 & 140 & 210 & 610 \\
${\rm \beta_{max} [km]}$               & 1.5$\rightarrow$150 & 580 & 19 000 & 64 000 \\
ff demag., $M \equiv \sqrt{\beta_{\rm max}/\beta^*}$
                                        & 220$\rightarrow$7100 & 17 000
                                        & 89 000 & 360 000 \\
chrom. quality factor, $Q \equiv M \cdot \delta$
                                       & 0.007$\rightarrow$11 & 10 & 10 & 45 \\
\hline
\multicolumn{1}{|l|}{\bf neutrino radiation parameters:} & & & & \\
collider reference depth, D[m]     & 10$\rightarrow$300 & 100 & 100 & 100 \\
ave. rad. dose in plane [mSv/yr]   & $2 \times 10^{-5}$$\rightarrow$0.02
                                        & 2.3 & 10 & 20 \\
str. sec. len. for 10x ave. rad. [m]
                                   & 1.3$\rightarrow$2.2 & 1.1 & 1.0 & 4.2 \\
$\nu$ beam distance to surface [km]    & 11$\rightarrow$62 & 36 & 36 & 36 \\
$\nu$ beam radius at surface [m]       & 4.4$\rightarrow$24 & 0.8 & 0.08 & 0.08

\end{tabular}
\label{colliderpara}
\end{table}

\subsection{Generation of the Parameter Sets}
\label{subsec:collider_gen}

  The parameter sets in table~\ref{colliderpara} were generated through
iterative runs of a stand-alone computer program, as has been described
previously~\cite{epac98, pac99he}.

  The most important physics parameter for a specified collider
energy is the luminosity, $\lum$. This is derived in terms of several
input parameters according to the formula~\cite{epac98}:
\begin{eqnarray}
\cal{L}
{\rm   [cm^{-2}.s^{-1}] }
   & = & {\rm 2.11 \times 10^{33} \times H_B
                   \times (1-e^{-2t_D[\gamma \tau_\mu]}) }
                                                           \nonumber \\
   & &  \times {\rm  \frac{ f_b[s^{-1}] (N_0[10^{12}])^2 (E_{CoM}[TeV])^3}
                          { C[km] } } \nonumber \\
   & & {\rm \times \left( \frac{\sigma_\theta [mr].\delta[10^{-3}]}
                   {\epsilon_{6N}[10^{-12}]}  \right) ^{2/3}  },
                                    \label{lum}
\end{eqnarray}
where the input variables are
the CoM energy (${\rm E_{CoM}}$), the collider ring
circumference (C), the beams' fractional momentum spread ($\delta$)
and 6-dimensional invariant emittance (${\rm \epsilon_{6N}}$), the
time until the beams are dumped (${\rm t_D}$),
the bunch repetition frequency (${\rm f_b}$),
the initial number of muons per bunch ($N_0$),
and the beam divergence at the interaction point ($\sigma_\theta$).
Units in equations throughout this paper are given in square brackets.
(The time-to-dump, $t_D$, is given in units of the boosted muon lifetime,
$\gamma \tau_\mu$.)
This formula uses the standard assumption from the Muon Collider
Collaboration that the ratio of transverse to longitudinal emittances
can be manipulated freely in the muon cooling channel to maximize the
luminosity for
a given ${\rm \epsilon_{6N}}$. The pinch enhancement factor, ${\rm H_B}$,
is very close to unity (see table 1), and the numerical coefficient in
equation~\ref{lum} includes a geometric correction factor of 0.76 for the
non-zero bunch length, $\sigma_z = \beta^*$ (the ``hourglass effect'') .

  In practice, the muon beam power and current are limiting parameters for
energy frontier muon colliders, so the parameters are actually chosen
to optimize the ``specific luminosity'':
\begin{equation}
l \equiv \frac{\cal{L}}{f_b \times N_0}.  \label{speclum}
\end{equation}
The luminosity is then determined from the choice of beam
current that corresponds
to the highest plausible beam powers.

  Several further parameters in table~\ref{colliderpara}
have been derived from the input parameters
that determine the luminosity. These include, for example, the beam-beam tune
disruption parameter, $\Delta \nu$. Other output parameters require
additional modeling assumptions and/or further input
parameters~\cite{epac98, pac99he}. Examples include some of the output parameters
for the final focus; these require both the input of a reference pole-tip
magnetic field for the final focus quadrupoles (${\rm B_{5\sigma}}$) and
a much simplified model for the final focus magnet lattice that is
a linearized extrapolation from existing final focus lattice designs
for lower energy muon colliders.

The physics parameters in table~\ref{colliderpara} include two examples of
event sample sizes. As is discussed in references~\cite{pr98,hemc99intro}
these give an indication of the physics potential corresponding to the
specified luminosity and energy. Briefly, the number
of $\mu\mu \rightarrow {\rm ee}$ events gives a benchmark estimate of the
discovery potential for elementary particles at the full CoM energy of the
collider, while the production of hypothesized 100 GeV Higgs particles
indicates roughly how the colliders might perform in studying physics
at a lower energy scale.

\subsection{Optimization of the 10 TeV and 100 TeV Parameter Sets}
\label{subsec:collider:opt}


\subsubsection{The Initial Choice of Energies}

  The two energies for the parameter sets, $\ECoM=10$ TeV and 100 TeV,
were chosen because they bracket that energy decade. The 10 TeV lower limit
was chosen to be well above the highest energy that had been studied in
detail, namely, $\ECoM=4$ TeV for the Snowmass'96 workshop~\cite{Snowmass}.
Further, the neutrino radiation for very high luminosity $\mm$ colliders
at 10 TeV and above is high enough to rule out siting them at an
existing laboratory, as is covered elsewhere in these
proceedings~\cite{hemc99nurad}. This necessitates a fresh
outlook for the design optimization of the $\mm$ colliders that is free
from site-specific preconceptions involving existing laboratories,
which was considered a good thing.

  The choice of the upper energy limit was more technically constrained.
For the 100 TeV parameter sets, the synchrotron radiation power had
risen to become almost identical to the beam power, signaling a clear upper 
bound for the feasibility of circular $\mm$ colliders.

  To preview later discussion, it is noted that our understanding
of the constraints on high energy muon colliders advanced during the workshop,
as will be covered in section~\ref{sec:assess}.
An additional constraint on the maximum
possible energies for circular muon colliders was
discovered~\cite{Telnovsynch},
due to beam heating arising from
the quantum mechanical nature of the synchrotron radiation.
On the other hand, the future prospects of many-TeV muon colliders
were given a boost when the possible potential for linear
colliders at even higher energies was uncovered~\cite{Zimmermann}.

\subsubsection{Balancing Luminosity against Technical Difficulty}

  After deciding on the collision energies, it was then decided that the
10 TeV (set A) and the first of the 100 TeV parameter sets (set B) should
assume only evolutionary changes in technology from the base-line parameters
that have been previously posited for lower
energy colliders~\cite{status}. For example,
the assumed 6-dimensional emittances are factors of 3.5 (10 TeV)
or 50 (100 TeV) smaller than the value $170 \times 10^{-12}\;{\rm m}^3$
that is normally used in Muon Collider Collaboration scenarios for first
generation muon colliders. The smaller emittances assume that the performance
of the muon cooling channel will be progressively
improved through further design optimization, stronger magnets,
higher gradient rf cavities and other technological advancements
and innovations.

  The second parameter set at 100 TeV (i.e., set C) encouraged study
on some of the possibilities for using exotic technologies to improve
the potential performance of future many-TeV $\mm$ colliders. The additional
assumed advances increased the luminosity by two orders of magnitude over
the evolutionary parameter set at 100 TeV, to
what would be a very impressive $1 \times 10^{38}\: \Lunits$. (The luminosity
should ideally rise as $\ECoM^2$, as is explained in
reference~\cite{hemc99intro}.) The hypothesized technical advances
included:
\begin{enumerate}
   \item  exotic cooling, to obtain a phase space density that is
a further 3 orders of magnitude larger than the assumption for the
evolutionary parameter set at 100 TeV
   \item  charge compensation at the interaction point (ip), to reduce
the effective charge by a factor of 10. This assumption led rather directly
to a corresponding increase in the luminosity by about a factor of 10.
   \item  more aggressive final focus parameters were included to allow
for potential improvements in the final focus design, perhaps using
exotic focusing technologies
   \item  the beam power was almost doubled from the evolutionary
parameter set (B), to ``top up'' the luminosity to
$1 \times 10^{38}\: \Lunits$ .
\end{enumerate}

\subsubsection{Final Focus Constraints}

  The final focus design may well present the most difficult
design challenges that are relatively specific to high energy
muon colliders. (This is to be contrasted with the muon cooling channel,
which is a formidable challenge for all muon colliders.)
References~\cite{epac98} and~\cite{pac99he} have previously addressed
the general design constraints and issues for final focus designs at
many-TeV muon colliders.

  To re-cap the discussion of references~\cite{epac98} and~\cite{pac99he},
higher energies demand progressively stronger focusing
to generate the smaller spot sizes needed to increase the luminosity.
Two simply defined parameters were used as benchmarks to obtain
final focus specifications that might provide plausible starting
assumptions for first attempts at magnet lattice designs.
Firstly, an overall beam demagnification parameter is
defined~\cite{epac98} in terms of one of the Courant-Snyder lattice
parameters, $\beta$, as
\begin{equation}
M \equiv \sqrt{  \frac{\beta_{\rm max}}{\beta^*} }.
   \label{M}
\end{equation}
This is a dimensionless parameter that gauges the strength of the
focusing. The size of $M$ should be closely correlated with
fractional tolerances in magnet uniformity, residual chromaticity,
etc., where the chromaticity is a measure of the change in response
of the final focus to off-momentum particles. Secondly, a high residual
chromaticity can be compensated for by decreasing the fractional momentum
spread of the beams, $\delta$. This suggests that another measure of
the final focus difficulty might come from the product of the
demagnification and momentum spread,
\begin{equation}
q \equiv M \delta,
  \label{q}
\end{equation}
where $q$ has been referred to~\cite{epac98} as the
``chromaticity quality factor''.

  In generating the parameter sets, the values of $M$ and $q$ were
compared to those for existing $\ee$ and $\mm$ final focus designs,
as was discussed in references~\cite{epac98,pac99he}. In practice, slightly
more attention was paid to $q$ than to $M$ in obtaining the final
parameters. It can be seen from table~\ref{colliderpara}
that the two ``evolutionary''
parameter sets, A and B, were constrained to the value $q=10$, which is very
similar to the calculated value, $q=11$, for the final focus lattice
design of the 3 TeV $\mm$ collider in reference~\cite{status}. A more
aggressive value, $q=45$, was allowed for in the second parameter set
at 100 TeV. 

  It is noted that the parameter sets at these high energies are always
limited by $\Delta \nu$ and it is useful and straightforward to rewrite
equations~\ref{lum} and~\ref{speclum} in the form:
\begin{equation}
l \propto \frac{\Delta \nu}{\beta^*},
   \label{Keileq}
\end{equation}
which has no explicit dependence on emittance or bunch size for a given
energy. The experience with optimizing the parameter sets was that this
independence is true as an approximation only~\cite{pac99he}; residual
dependences on limiting magnet
apertures etc. meant that, in practice, it was almost always
possible to slightly improve the specific luminosity by
re-optimizing to parameter sets with smaller assumed emittances.

  A value of $\Delta \nu = 0.10$ was assumed for all parameter sets.
This was estimated by interpolating the results from a beam tracking study
described in reference~\cite{Snowmass}. Equation~\ref{Keileq} and the
discussion that follows indicate that the luminosity will scale approximately
linearly with different assumed values for $\Delta \nu$.

\subsubsection{Constraints on Energy Spread from Beamstrahlung}

  The ``chromaticity quality factor'' figure of merit, equation~\ref{q},
favors decreasing the fractional momentum spread, $\delta$, in order to
ease the difficulty of the final focus, and this strategy was found to be
effective in optimizing the luminosity for all three parameter sets.
 By the $\ECoM=100$ TeV energy scale, however, the value of $\delta$
was found~\cite{pac99he} to be limited from below by the rapidly rising
beamstrahlung at collision. This occurred even though the fractional
beamstrahlung energy loss, $(\Delta E)_{brem}$,
remained at the level of parts-per-million per beam
crossing, i.e., much less than the percent level expected at TeV-scale
linear $\ee$ colliders. The difference is the need for multiple passes
at $\mm$ colliders, which compounds the sensitivity to beamstrahlung
losses.

  The average beamstrahlung energy losses can be replaced by rf
acceleration, of course. However, the particle-by-particle variations
will contribute to the spread in the beam momentum, and any such
contributions from beamstrahlung must be limited to somewhat below the
original momentum spread of the beam. The residual
contributions to the beam energy spread should rise as the square root
of the number of passes, since they will be statistically independent
from turn to turn. Therefore, an appropriate criterion that was chosen
to set lower limits on $\delta$ is:
\begin{equation}
\frac{ (\Delta E)_{brem} \times \sqrt{n^{eff}_{turn}} }{\delta}
                     \stackrel{<}{\sim} 1,
   \label{bremlim}
\end{equation}
where the effective (i.e. luminosity-weighted) number of turns,
$n^{eff}_{turn}$, has values in the range $n^{eff}_{turn}\simeq 1000$.
The evolutionary (B) and ultra-cool (C) parameter sets
at $\ECoM=100$ TeV had chosen
values of 0.49 and 0.33 for the left hand side of equation~\ref{bremlim},
respectively.

  As an aside, it is noted that reference~\cite{pac99he} had suggested
following the lead of proposed TeV-scale $\ee$ colliders by considering
the option of using flat, rather than round, beam spots at the ip 
in order to reduce the beamstrahlung. This was tried, but all attempts
led to disappointing luminosities and so round beam spots were retained
for the parameter sets.

\section{Straw-man Acceleration Parameters}
\label{sec:accel}

\begin{table}[htb!]
\caption{
Straw-man acceleration parameter sets for high energy
muon colliders.
The word ``net'' in the column ``net ${\rm E_{rf}}$''
refers to the net energy gain per turn in the rf cavities
after approximately subtracting synchrotron radiation
losses from the 50 GeV and 250 GeV energy gains in
the first and second recirculators, respectively.
The parameter sets
${\rm N_f^A}$,
${\rm N_f^B}$ and
${\rm N_f^C}$
are the numbers of muons per bunch at the exit of each FFAG
corresponding to each of the three straw-man muon collider ring
scenarios in table~\ref{colliderpara}.
}
\begin{center}
\begin{tabular}{|ccccccccccc|}
\hline
$E_i$ & $E_f$ & $\frac{E_f}{E_i}$  & circum. & ${\rm B_{ave}}$ &
    net ${\rm E_{rf}}$ & \# turns &$f_{\rm decay}$                      &
    ${\rm N_f^A}$ &  ${\rm N_f^B}$ &  ${\rm N_f^C}$ \\
 ${\rm [TeV]}$ & [TeV] &                    &  [km]   &       [T]       & 
             [GeV]     &         &                         \%     &
             [$10^{12}$] & [$10^{12}$] & [$10^{12}$] \\
\hline
     &  0.5  &      &      &       &     &       &       &3.35&1.038 &0.247 \\
0.50 & 1.25  & 2.5  &  15  &  1.7  &  50 &  15   & 4.3\% &3.21&0.993 &0.236\\
1.25 & 2.50  & 2.0  &  15  &  3.5  &  50 &  25   & 3.3\% &3.10&0.961 &0.229\\
2.50 & 3.50  & 1.40 &  15  &  4.9  &  50 &  20   & 1.6\% &3.05&0.945 &0.225\\
3.50 & 4.55  & 1.30 &  15  &  6.4  &  50 &  21   & 1.3\% &3.01&0.933 &0.222\\
4.55 & 5.00  & 1.10 &  15  &  7.0  &  50 &   9   & 0.5\% &3.00&0.929 &0.221\\
 5.0 & 12.5  & 2.5  & 100  &  2.6  & 250 &  30   & 5.7\% &    &0.876 &0.208\\
12.5 & 25.0  & 2.0  & 100  &  5.2  & 249 &  50   & 4.4\% &    &0.838 &0.199\\
25.0 & 35.0  & 1.40 & 100  &  7.3  & 246 &  41   & 2.2\% &    &0.820 &0.195\\
35.0 & 45.5  & 1.30 & 100  &  9.5  & 238 &  44   & 1.8\% &    &0.805 &0.192\\
45.5 & 50.0  & 1.10 & 100  & 10.5  & 229 &  20   & 0.7\% &    &0.800 &0.190\\
\end{tabular}
\label{accel}
\end{center}
\end{table}

\begin{figure}[t!] %
\centering
\includegraphics[height=3.0in,width=4.0in]{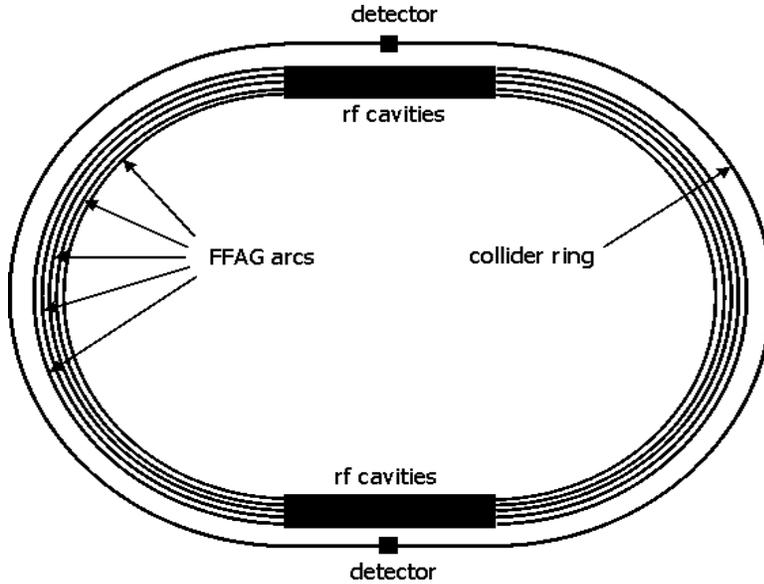}
\caption{
Accelerator layout for the acceleration scenario of table~\ref{accel}.
The layout is schematic and is certainly not drawn to scale.
A single tunnel contains
5 rings of FFAG arcs. All the arcs pass through the same
rf cavities, shown here in 2 linacs on opposite sides
of the tunnel. The collider ring is also shown in the same
tunnel, indicating that this accelerator complex brings
the beam up to collision energy, i.e., this could be the
0.5-5 TeV ring for the 10 TeV collider (parameter set A)
or the 5-50 TeV ring for the 100 TeV colliders (sets B and C).
Transfer lines between the rings are not shown.
As an aside, 2 detectors are shown in the collider storage ring,
although this was not assumed in the
workshop. This would double the luminosity but would complicate
the design of the storage ring.
}
\label{acclayout}
\end{figure}

\subsection{Introduction}
\label{subsec:accel_intro}


 Table~\ref{accel} gives straw-man acceleration scenarios that reproduce the
final energy and bunch charge for each of the three straw-man muon collider
ring scenarios given in table~\ref{colliderpara},
labeled as A) 10 TeV with $10^{36}$
luminosity, B) 100 TeV with $10^{36}$ luminosity and C) 100 TeV with
$10^{38}$ luminosity. The layout of each of the two recirculating complexes
for table~\ref{accel} is sketched schematically in figure~\ref{acclayout}.

  The acceleration scenarios of table~\ref{accel} and figure~\ref{acclayout}
will be described in subsection~\ref{subsec:accel_scenario}. For now, we note
that the table contains only a minimal amount of information -- much less
than was provided for the collider ring -- and, in practice, the acceleration
parameters were much less critical than the collider ring parameters for
determining the technical feasibility or otherwise of the collider scenarios.
This viewpoint is supported by a much more detailed and knowledgeable
acceleration scenario that is presented elsewhere in these
proceedings~\cite{Berg}.

  Aside from the technical considerations, the acceleration is expected to
dominate the cost of the colliders so its cost optimization will be very
important and this was the main design criterion for the straw-man scenarios
presented in table~\ref{accel}. To minimize the cost, the scenarios use
configurations of recirculating linacs with ``fixed field alternating
gradient'' (FFAG) magnet lattices.

  The rest of this section is organized as follows. A very simplistic and
non-technical introduction to FFAGs will be given in the next subsection.
Some preliminaries on calculating decay losses during acceleration occupy
the subsection after that before, in subsection~\ref{subsec:accel_scenario},
returning to describe the motivation for the parameter choices in
table~\ref{accel}.

\subsection{FFAG Recirculating Arcs}
\label{subsec:accel_FFAG}


\begin{figure}[t!] %
\centering
\includegraphics[height=2.5in,width=5.5in]{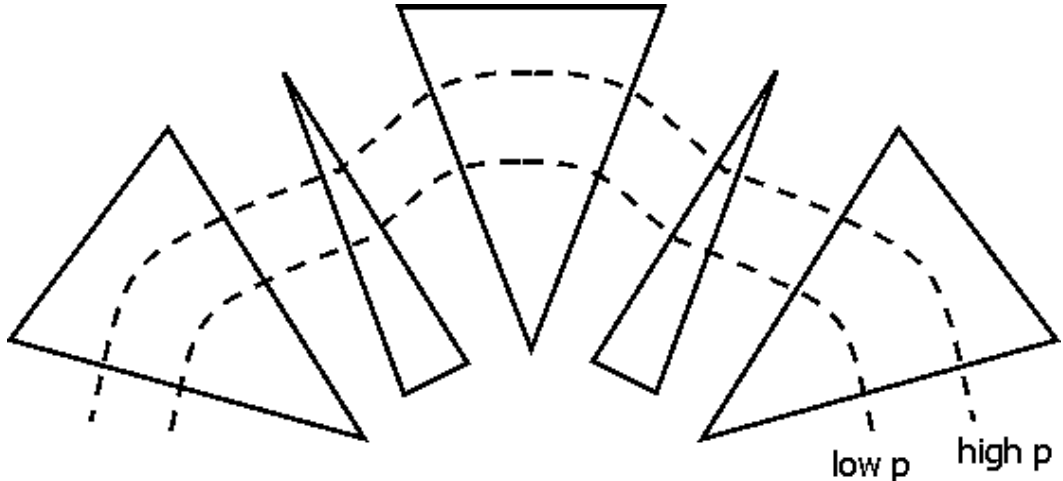}
\caption{
A very schematic illustration of the FFAG concept. Each triangle
signifies a bending magnet with a non-uniform magnetic field. The acute
point of each triangle signifies the direction of the bending magnetic
field and the thickness at any radius signifies the magnetic field
strength in this direction rather than the spatial extent of the magnet.
(More generally, the increase in the field gradient will not be linear.)
If the magnet spacings and magnetic field parameters are appropriate then
the non-uniform bending fields automatically provide alternating gradient
focusing in both transverse planes.
}
\label{FFAGconcept}
\end{figure}

  The amount of expensive rf acceleration can be reduced many-fold
relative to linear
accelerators by bending the muons around for many passes through the
same length of linac. The onus then shifts to minimizing the cost of the
magnets in the recirculating arcs. In turn, it is then desirable that each
of the arcs be able to accept a wide range of momenta so it can be reusable
for many traverses. The most promising option for doing this appears to lie
in a class of either quadrupole-loaded or combined function magnet lattices
that are referred to as ``fixed focus, alternating gradient'' or ``FFAG''
lattices. Fast-ramping synchrotrons may also be
considered~\cite{Snowmass,status} but steady-state operation of FFAGs
appears likely  to be cheaper and might well be more reliable.

  Figure~\ref{FFAGconcept} gives a very conceptual illustration of the
basic idea of FFAGs. It can be seen that the alternating sign of the
bending field results in net bending in the direction of the stronger
dipoles a provides the scalloped beam trajectories that are characteristic
of FFAGs. Further out trajectories see a progressively
stronger average bending field and so are appropriate for transporting
larger momenta in proportion to the average magnetic field strength.

  FFAGs were first considered back in the 1950's~\cite{FFAGhistory}
but, presumably, were not developed further at that time because the
simpler alternative of slowly ramping synchrotrons was adequate for
the acceleration of stable particles. Impressively, FFAG lattices
have now been designed that transport as much as factors of 5 to 10
in muon momentum, although such extreme designs require very large
apertures and the peak magnetic fields are several times the average
bending field. Some initial design studies for more practical FFAG
lattices are presented elsewhere in these proceedings~\cite{Garren, Dejan}.

  Perhaps the biggest technical problem with all FFAG scenarios is
the difficulty in maintaining turn-by-turn an appropriate phase
relationship with the rf acceleration, since the path lengths
of the muon orbits within the FFAG lattice get progressively larger
with increasing energy -- as is conceptually illustrated in
figure~\ref{FFAGconcept}.
It is a nice feature of many-TeV colliders that these
problems become progressively less at higher energies because
the increasing revolution period through the arcs gives more
time for adjustments between passes through the linac.

\subsection{Rf Acceleration and Decay Losses}
\label{subsec:accel_decay}

 The amount of radio-frequency (rf) acceleration per turn will be
determined by a trade-off between minimizing the expense and
the tunnel length occupied by rf (favors less rf) and minimizing
the number of turns and the decay losses (favors more rf).

 A formula relating the decay losses to the rf and recirculator
parameters can be derived directly from the decay equation for
the change in the number of muons, $N$, with distance, $x$:
\begin{equation}
\frac{-1}{N} \frac{dN}{dx} = \frac{1}{\beta \gamma c \tau},
\end{equation}
where $c$ is the speed of light, the scaled muon velocity
is essentially unity
for the muon energies under consideration, $\beta=1$,
$\gamma \equiv \frac{E_\mu}{m_\mu c^2}$ is the conventional
relativistic gamma factor and the muon
mass and its lifetime, $\tau$, are such that
$\frac{mc^2}{c \tau}=0.1604\; {\rm GeV.km^{-1}}$.

 It follows easily that muon decay losses lead to ratios
of initial to final bunch populations, $\frac{N_i}{N_f}$,
that are related to the recirculator tunnel lengths in
units of kilometers, $L^j[km]$, the number of GeV per turn
of rf acceleration, $E_{rf}^j[GeV]$, and the ratio of
final to initial energies in the recirculator, 
$\frac{E_i^j}{E_f^j}$,  through
\begin{equation}
\ln \left( \frac{N_f}{N_i} \right) =
     0.1604 \sum_{j=1,N} \frac{L^j[km]}{E_{rf}^j[GeV]}
          \ln \left( \frac{E_i^j}{E_f^j}  \right),
       \label{accloss}
\end{equation}
where $j=1,N$ is the index for the $j^{th}$ of N
recirculators. Equation~\ref{accloss} has made the approximation of
averaging the acceleration to an assumed constant gradient over
the length of the recirculator
rather than the real situation where it will be concentrated in
one or more rf linacs placed around the recirculator. This should
introduce only small fractional errors in the calculated particle
losses for the parameters given in table~\ref{accel}.

\subsection{Optimization of the Straw-man Acceleration Scenario}
\label{subsec:accel_scenario}


   The straw-man acceleration scenario presented
in table~\ref{accel} starts at 500 GeV, working
on the assumption that the acceleration to this energy range has already
been developed and used for a previous TeV-scale $\mm$ collider. The
acceleration scenario for the $\ECoM=10$ TeV collider (set A) then needs
to provide
exactly one decade of energy gain, accelerating the beams from 0.5 TeV up
to their collision energy of 5 TeV. It economizes on expensive rf
acceleration by utilizing only a relatively modest 50 GV of rf cavities.

  The $\ECoM=100$ TeV collider scenarios start with the $\ECoM=10$ TeV
acceleration scenario and add a further decade of acceleration to raise
the beam energies to 50 TeV. A further 250 GV of rf cavities are utilized,
which is, for example, much less rf than is required for the next generation
of $\ee$ colliders and so should be easily compatible with the budget
constraints on a 100 TeV collider.

  In more detail, it can be seen from table~\ref{accel} that the
recirculating accelerator to 50 TeV is essentially a scaled copy of
that to 5 TeV. Both recirculators use 5 rings of FFAG arcs and the
fractional momentum increment in each of the 5 rings is the same
between the first and second recirculator. As one difference,
the average bending fields, ${\rm B_{ave}}$, in the second recirculator
are assumed to be a factor of 1.5 times higher than in the first.

  Figure~\ref{acclayout} is a schematic diagram for a possible layout of
either of the two recirculators. As a specific suggestion of this layout,
it assumes the 5 FFAG rings to be housed in the same tunnel. Further,
this tunnel is assumed to be the collider tunnel, i.e., the 5+5 TeV collider
for the first recirculator and the 50+50 TeV collider for the second.

  The obvious motivation for the layout of figure~\ref{acclayout} is to
minimize the complexity and tunnel expense of the scenario. However,
requiring all 5 FFAG rings in a recirculator to have the same radius
as the collider ring has the obvious consequences of fixing the FFAG ring
radii and of constraining the average bending magnetic fields in each ring
according to the ranges of transported momenta in that ring.
Table~\ref{accel} gives a specific scenario for doing this.

  The design of
the later FFAG rings in each recirculator is clearly more constrained
than those for the earlier rings because the average bending field
must be closer to the (assumed high) average bending field of the
collider ring. This is dealt with in table~\ref{accel} by constraining
the momentum swing to become progressively smaller for the later rings
in the recirculators.
Assumed energy ranges covered by the arcs range from a factor
of 2.5 increase -- for the lowest energy arcs in each recirculator --
down to 10\% energy gain for the highest energy arcs in each
recirculator. These are really no more than guesses since, for
example, the magnet apertures and ratios of peak-to-average
magnetic fields required for this scenario are unknown.

  Assumed average gradients for superconducting rf of 25 MV/m,
as is assumed for the proposed TESLA $\ee$ collider,
correspond to total rf lengths of 2 km (10 km) for the 10 TeV (100 TeV)
colliders, which is 13.3\% (10.0\%) of the collider ring circumference.
The example schematic layout of figure~\ref{acclayout} shows the rf to be
split equally between the two straight sections of tunnel on the opposing
sides of the ``race-track'' collider ring, although this choice was
somewhat arbitrary.

  Decay losses were calculated according to equation~\ref{accloss}.
Non-decay losses
were neglected. Synchrotron radiation energy losses -- which range up to
about 10\% per turn at 50 TeV -- have been included in a simple approximate
manner. Table~\ref{accel} shows the overall decay losses to be acceptably low,
at 10.5\% and 13.9\% respectively, for each of the two decades of energy gain.

  Having detailed the scenario, it should again be emphasized that the
overall scenario, together with its specific choices and assumptions,
was intended to do no more than provide the seed for more credible design
studies from the accelerator physicists attending this workshop. Nothing
but the qualitative assumptions of the scenario should be considered at all,
and even these only at the reader's discretion. Of course, none of the specific
numerical assumptions should be taken at all seriously, beyond perhaps
obtaining a rough qualitative feel for such parameters as the amount of rf
acceleration required and the magnitudes for the fractional decay losses.

  Bearing the preceding paragraph in mind, we conclude this section by again
referring the reader to the vastly more competent and detailed acceleration
studies that emerged from the workshop: the overall acceleration
scenarios of reference~\cite{Berg} and the FFAG design studies in
references~\cite{Garren} and~\cite{Dejan}.

\section{Assessment of the Muon Collider Parameter Sets at HEMC'99}
\label{sec:assess}

\begin{table}[htb!]
\caption{An assessment of the feasibility of high energy collider parameter
sets, incorporating the advances in understanding from the HEMC'99
workshop. See text for details.
}
\begin{center}
\begin{tabular}{|r|ccc|}
\hline
\multicolumn{1}{|c|}{ {\bf parameter set} }
                            & A  & B  & C \\
\multicolumn{1}{|c|}{ {\bf center of mass energy, ${\rm E_{CoM}}$} }
                            & 10 TeV  & 100 TeV  &  100 TeV \\
\multicolumn{1}{|c|}{ {\bf additional description} }
                             & evol. extrap. & evol. extrap. & ultracold beam, etc. \\
\hline \hline
\multicolumn{1}{|l|}{\bf Luminosity for Physics:}
                                & excellent     &     fair   & excellent\\
                         &             &               &                \\
\multicolumn{1}{|l|}{\bf Technology:} &&& \\
acceleration             & probably OK &           OK  &             OK \\
detector backgrounds     & probably OK  & probably OK  & probably OK    \\
beam cooling             & probably OK & probably OK   & problematic    \\
synch. radiation         & probably OK & borderline    & NOT FEASIBLE   \\
final focus              & challenging & problematic   & problematic    \\
{\bf overall technology:}& challenging & problematic   & NOT FEASIBLE   \\
                         &             &               &                \\
\multicolumn{1}{|l|}{\bf Cost:}
                         & challenging & problematic & problematic \\
                         &             &               &                \\
neutrino rad./siting     & dedicated new site & same site & same site    \\
                         &             &               &                \\
{\bf OVERALL}            & challenging & problematic   &   NOT FEASIBLE \\
\end{tabular}
\label{assess}
\end{center}
\end{table}

 This section reviews the studies and assessments at HEMC'99 of the
collider ring and acceleration parameter sets of
tables~\ref{colliderpara} and~\ref{accel}.
It will concentrate on the muon collider design issues arising
out of the parameter sets.
The reader is also referred to the summary paper by Willis~\cite{Willis}
for a more general overview of the findings of the workshop.

   Table~\ref{assess} summarizes the status of the acceleration
and collider parameter sets after review at the workshop.
As an important piece of contextual information,
the assessment of parameter set A (10 TeV), assumes that a
TeV-scale muon collider has
already been built and successfully operated and the parameter set
in each successive column assumes that the collider of
the preceding column has already been built.

  The following subsections have been grouped according to
subject areas that follow fairly closely, but not exactly, the
rows of table~\ref{assess}: on luminosity, acceleration, detectors,
cooling, synchrotron radiation, final focus design and beam
instabilities. A more general outlook and list of conclusions based
on these observations will be deferred to the final section,
section~\ref{sec:concl}.

\subsection{Assessment of Luminosities for Physics}
\label{subsec:assess_lum}

  The luminosity requirements for $\mm$ colliders are discussed
in some detail elsewhere in these proceedings~\cite{hemc99intro}.
Ideally, collider luminosities should rise as $\ECoM^2$ and it
is seen that, indeed, the luminosities for all three parameter sets
in table~\ref{colliderpara} are higher than for any existing or
(to the author's knowledge) other proposed collider.

  Both parameter sets A and C have excellent luminosities, even
considering their high energies, while the luminosity of parameter
set B was still considered  to be ``fair'' for a 100 TeV lepton collider.
(See reference~\cite{hemc99intro} for further discussion.)

\subsection{Assessment of the Acceleration Scenario}
\label{subsec:assess_acc}

  Studies at HEMC'99 focused more on the collider ring parameter
sets of table~\ref{colliderpara} than on the
acceleration scenario of table~\ref{accel}.
The acceleration scenario was considered critical mostly to the extent
that it would be expected to be the biggest single component of the
overall cost of the collider. Unfortunately, the cost of the FFAG magnets
was not able to be explicitly addressed in any detail due to the newness
and developing nature of FFAG scenarios~\cite{Garren, Dejan} for
muon colliders. A rather indirect source for some optimism on the
acceleration costs could come from any assumed correlation with some
relatively
favorable cost estimates for the collider ring magnets, by
Harrison~\cite{Harrison}, who roughly assessed the cost for the
collider magnets for the 10 TeV scenario (set A) to be perhaps of
order 400 million dollars.

  Technically, muon acceleration tends to get easier at higher
energies due to the increasing muon lifetime, smaller beam sizes
and lower circulation frequencies in recirculating linacs. Hence,
the technical feasibility of acceleration up to the energies in the
table is automatically established to a large extent by the assumed
previous success of the acceleration at a TeV-scale $\mm$ collider.
As a minor caveat to this, Harrison pointed out the increased
load due to synchrotron radiation in the FFAGs. However, the collider
ring magnets will need to handle the synchrotron radiation load for
many times more turns than the FFAG arcs, so even this technical difficulty
is concentrated more in the collider ring than the accelerating
lattice.

  Berg~\cite{Berg} pointed out that slightly increased technical difficulties
might instead be expected for the {\em low energy} end of the acceleration
for parameter sets A and, especially, B. This could result from the higher
specified values for the longitudinal emittance in the many-TeV parameter
sets: table~\ref{colliderpara} shows the longitudinal emittances for these
parameter sets to be, respectively, similar to, and about twice as large as,
the longitudinal emittance for the 3 TeV parameter set of
reference~\cite{status}.

\subsection{Detector Backgrounds}
\label{subsec:assess_det}

  All muon collider detectors face challenging backgrounds resulting
from the electron daughters of decaying muons near the interaction point.
However, the amount of electromagnetic ``junk'' entering the detector
is relatively independent of the collider energy since the power
density of deposited electromagnetic energy depends primarily on
the beam current rather than the beam energy. (For confirmation of
this statement, see the values in the ``power density into magnet
liner'' row of table~\ref{colliderpara}.) Hence, such backgrounds
are expected to be manageable for these many-TeV parameter sets
under the stated assumption that the problem has already been solved
at TeV-scale collider detectors. (A specific strategy for handling
these backgrounds that was developed at the workshop is described
in reference~\cite{Rehak} of these proceedings.)

  Muons entering the side of the detector, either from beam halo
or Bethe-Heitler $\mm$ pair production, are the one background
that is expected to evolve markedly with energy. As muons become
more relativistic they become less and less like minimum-ionizing
particles and deposit larger amounts of energy ``catastrophically''
in, mainly, electromagnetic showers. This issue was not addressed
at the workshop and it deserves further study.

\subsection{Beam Cooling}
\label{subsec:assess_cooling}


  Parameter sets A and B assume only evolutionary improvements
in the ionization cooling performance over that assumed (but far
from demonstrated~\cite{status}!) for TeV-scale colliders so,
by definition, the beam cooling should probably be OK if following
on from the TeV-scale collider.
Parameter set C is very different, assuming that some form
of exotic cooling will be able to increase the phase space density
of the muon beams by three orders of magnitude from that assumed
for parameter set B.

  Such ultra-cold muon beams are still looking plausible but have
not yet progressed beyond that. The most promising of the exotic
cooling methods is optical stochastic cooling~\cite{Zholents}.
This method clearly has formidable technical challenges but no
obvious show-stoppers. Other, very low energy, cooling methods
were also presented at the workshop~\cite{Lebrun, Nagamine}. There
is some concern that any cooling method using non-relativistic
muons (i.e. with scaled velocity $\beta \ll 1$) may well not be
feasible for preparing the high-charge muon bunches needed for
colliders, due to space charge limitations.

  It is noted that parameter set C provides a specific example
of a general feature for ultra-cold muon beams. Since the collisions
at many-TeV colliders would normally be tune-shift limited anyway,
it is likely that improved cooling would also require ip compensation
to substantially benefit the luminosity. We now discuss yet another
barrier to the use of ultra-cold beams, at least at very high
energies, from synchrotron radiation.

\subsection{Synchrotron Radiation}
\label{subsec:assess_synch}

  It has already been noted that the synchrotron radiation power
in the 100 TeV colliders is already comparable to the beam power.
During the workshop, Telnov~\cite{Telnovsynch} raised what might
possibly be a stronger constraint from synchrotron radiation on the
energy reach of circular muon colliders, namely, the quantum nature
of synchrotron radiation may lead to heating, rather than damping,
of the horizontal beam emittance if the beam energy is high enough
and the emittance is already very small.

  Telnov's observation clearly spells the end of parameter set C,
with its ultra-cold beam at $\ECoM=100$ TeV. The other parameter
set at 100 TeV (set B) is also borderline, with an initial
horizontal emittance that is larger by a factor of five~\cite{Telnovsynch}
than the equilibrium emittance due to this effect, as calculated
by Telnov using a simple approximate model.

  The most likely possible loop-hole for parameter set B is that the
heating effect is reduced for a very strongly focusing collider lattice.
More specifically, Telnov's equation 2 shows the equilibrium emittance
to be proportional to the average of the ``H-function'' around the collider
ring, where
\begin{equation}
 H \propto \frac{\beta^3}{\rho^2},
    \label{H}
\end{equation}
for $\beta$ the standard Courant-Snyder parameter and $\rho$ the
collider ring's radius. (Stronger focusing corresponds to smaller
$\beta$ values around the ring.)

  To consider adjustments to parameter set B, equation~\ref{H} suggests that
100 TeV colliders with the emittances expected from ionization cooling
still look to be feasible by increasing the ring radius, $\rho$,
by, for example, a factor of two. This would lower both the equilibrium
emittance by a factor of four and the radiated energy per turn by
a factor of two, which is substantial compensation for halving
the number of collisions per bunch.

  A much more dramatic approach to beating the energy limits from
synchrotron radiation has come from Zimmermann~\cite{Zimmermann},
in the form of single pass linear $\mm$ colliders. Example parameter
sets are included in Zimmermann's paper, and are commented on in
more detail elsewhere in these proceedings~\cite{hemc99intro}.

\subsection{Final Focus Design}
\label{subsec:assess_ff}

  The final focus design extrapolations discussed in
section~\ref{sec:collider}
seemed to work well for the 10 TeV parameter set A. A magnet layout
for the final focus from Johnstone~\cite{Carol} closely reproduced
the predicted $\beta_{max}$ in table~\ref{colliderpara}. Further,
the lattice design experts at the workshop seemed to appreciate the
extremely challenging nature of the 10 TeV final focus parameters
without everybody actually condemning them as being clearly unrealistic,
i.e., an appropriate level of difficulty for a workshop of this nature!
See reference~\cite{Zimmermann} for more detailed studies and comments.

  The 100 TeV parameter sets were less fortunate. Even the ``evolutionary''
parameter set B was immediately dismissed by the lattice
experts as being incompatible with any final focus lattice designs
using conventional magnets. It will be very useful to get further
feedback on what exactly broke down in the simplistic
energy extrapolation that was described
in section~\ref{subsec:collider_gen}. Hopefully, such
feedback can then be used to obtain a better parameterization of the
energy evolution in the final focus parameters.
A more realistic and better established parameterization could then
be used to predict the
luminosity scaling with energy that might be expected
using conventional final focus technologies.

  Finally, two exotic final focus options were discussed that might go
beyond conventional magnet designs: ``dynamic focusing'' (using auxiliary
beams to focus the colliding beams) and plasma focusing. Discouragingly,
both options looked much less plausible than when considered for
single pass $\ee$ colliders, due to both the need for multiple passes
and the larger bunch currents assumed for $\mm$ collider parameters.
Also disappointing are the obstacles to beam compensation at collision
(as was assumed in parameter set C), which call into question the
possibility of being able to do this -- see reference~\cite{Telnovcomp}
for discussion on this topic.

\subsection{Beam Instabilities in the Collider Ring}
\label{subsec:assess_instab}

  Papers by Keil~\cite{Keil} and Zimmermann~\cite{Zimmermann}
provide studies on beam instabilities. Keil provides a
systematic assessment of the classes of instabilities,
including parameter comparisons with the LHC collider ring.
Zimmermann's tracking studies demonstrated that even circulating
the beams for a single turn should not be taken for granted,
let alone for of order 1000 turns over the lifetime of the
muons.

  As a connection to the physics capabilities of the collider ring
that needs to be borne in mind, the common assumption~\cite{status}
of collider rings that are isochronous is disfavored for retaining the
beam polarization. (See also reference~\cite{Heusch} for a discussion
on the importance of polarization.) As a rough hand-waving explanation,
the rate of polarization precession while circulating in the collider
ring is proportional to the muon's energy. It is intuitively clear
that the polarization will decay away more slowly if the energies of
all the particles are allowed to slosh around the beam average energy --
sometimes gaining in polarization precession (higher energy) over the
bunch average precession and sometimes losing (lower energy than the
bunch average). This is what happens in a collider ring with longitudinal
focusing as opposed to isochronous rings. The same argument also
favors small beam energy spreads.

%
%

\section{Outlook and Conclusions}
\label{sec:concl}


   The preceding section has reviewed the insights from HEMC'99
on the parameter sets of tables~\ref{colliderpara} and~\ref{accel}.
More generally than this, HEMC'99 has provided the first speculative
insights into (i) the ultimate physics potential for future colliders
at the high energy frontier and (ii) the potential challenges to reaching
very high energies with muon colliders. A personal interpretation of the
workshop's findings through the energy decades is:
\begin{itemize}
  \item {\bf muon colliders to the TeV scale:}
(added for completeness -- these energies were not discussed
in detail at the workshop) beam cooling
is the dominant technical challenge. Other major challenges
are the final focus region, backgrounds in the detector,
cost-efficient acceleration and beam stability throughout the
cooling, acceleration and storage in the collider ring.
Neutrino radiation will impose
significant design constraints and the beam currents may
be well below those for the straw-man parameters in
reference~\cite{status}
(i.e. $6 \times 10^{20}$ muons/sign/year in collision).
  \item {\bf to advance to the 10 TeV scale:} neutrino
radiation will probably dictate a new site. The final
focus region of the collider and magnet cost reduction
for acceleration may be the other major technical
design issues.
  \item {\bf to advance to the 100 TeV scale:} major
breakthroughs are needed in magnet costs and in the
final focus region.
  \item {\bf to advance to the 1 PeV scale and beyond:} this is
not absolutely ruled out in the far distant future
using a linac and many technological breakthroughs,
as illustrated by the parameter set in reference~\cite{Zimmermann}
and discussed further in reference~\cite{hemc99intro}.
\end{itemize}

  It would certainly be very valuable to follow up on the
understandings gained at this workshop. As a small first step,
modified parameter sets for many-TeV muon colliders are being
generated~\cite{epac2000} that take into account the insights gained
at HEMC'99. As a refinement to make interpolations easier, a parameter
set at the intermediate center-of-mass energy of 30 TeV will be
included.

  More substantially, there is need for a new study and workshop.
Preferably, this should include all three of the main accelerator
technologies -- pp, $\ee$ and $\mm$ colliders. This is
motivated~\cite{hemc99intro} both for a more coherent understanding
of the future of experimental high energy physics and in recognition
that the three accelerator technologies are deeply intertwined.
Planning is underway for such a study to take place in the Summer
and Fall of 2001.


\begin{references}

\bibitem{hemc99nurad}  B.J. King,
   {\it Neutrino Radiation Challenges and Proposed Solutions for
   Many-TeV Muon Colliders}, these proceedings.
   Also available from~\hfill\break
   \verb|http://pubweb.bnl.gov/people/bking/heshop/hemc_papers.html|.
\bibitem{epac98}   B.J. King,
   {\it Discussion on Muon Collider Parameters at Center of Mass
     Energies from 0.1 TeV to 100 TeV}, Proc. EPAC'98, BNL--65716.
    available from LANL preprint archive as {\it physics/9908016}.
\bibitem{pac99he}   B.J. King,
     {\it Muon Colliders from 10 TeV to 100 TeV},
     Proc. PAC'99, New York, 1999, pp. 3038-40,
     available from LANL preprint archive as {\it physics/9908018}.
\bibitem{pr98}   B.J. King,
     {Muon Colliders: New Prospects for Precision Physics and the
      High Energy Frontier},
     Proc. Second Latin American Symposium on High Energy Physics,
      San Juan, Puerto Rico, 8-11 April, 1998,
     available from LANL preprint archive as {\it hep-ex/9908041}.
\bibitem{hemc99intro}
      B.J. King,
    {\it  Prospects for Colliders and Collider Physics to the
          1 PeV Energy Scale},
          these proceedings.
          Also available from~\hfill\break
          \verb|http://pubweb.bnl.gov/people/bking/heshop/hemc_papers.html|.
\bibitem{Snowmass}
    The Muon Collider Collaboration,
    {\it $\mm$ Collider: A Feasibility Study},
    BNL-52503, Fermilab-Conf-96/092, LBNL-38946, July 1996.
\bibitem{Telnovsynch}   Valery Telnov,
      {\it Limit on Horizontal Emittance in High Energy Muon Colliders
             due to Synchrotron Radiation},
          these proceedings. Also available from~\hfill\break 
          \verb|http://pubweb.bnl.gov/people/bking/heshop/hemc_papers.html|.
\bibitem{Zimmermann}  F. Zimmermann,
    {\it Final Focus Challenges for Muon Colliders at Highest Energies},
            {\it ibid.}
\bibitem{status}
      The Muon Collider Collaboration,
      {\it Status of Muon Collider Research
      and Development and Future Plans},
      Phys. Rev. ST Accel. Beams, 3 August, 1999.
\bibitem{Berg}   J. Scott Berg,
         {\it Acceleration for a High Energy Muon Collider},
          these proceedings. Also available from
          \verb|http://pubweb.bnl.gov/people/bking/heshop/hemc_papers.html|.
\bibitem{FFAGhistory}   K.R. Symon,
         {\it The FFAG Synchrotron -- MARK 1}, MURA-KRS-6,
          November 12, 1954, pp. 1-19.
\bibitem{Garren}   Al Garren,
         {\it A Scaling Radial-Sector FFAG Lattice for a Muon Accelerator},
          these proceedings. Also available from~\hfill\break 
          \verb|http://pubweb.bnl.gov/people/bking/heshop/hemc_papers.html|.
\bibitem{Dejan}   Dejan Trbojevic, Ernest D. Courant and Al Garren,
         {\it FFAG Lattice Without Opposite Bends}, {\it ibid.}
\bibitem{Willis}   Bill Willis,
         {\it Muon Collider Workshop Summary}, {\it ibid.}
\bibitem{Harrison}  Mike Harrison,
          {\it Magnet Challenges: Technology and Affordability},
          oral presentation at this workshop. Transparency copies can
          be viewed at~\hfill\break
          \verb|http://pubweb.bnl.gov/people/bking/heshop/hemc_papers.html|.
\bibitem{Rehak}    P. Rehak {\em et al.},
        {\it Detector Challenges for $\mm$ Colliders in the 10-100 TeV
             Range},
          these proceedings. Also available from~\hfill\break
          \verb|http://pubweb.bnl.gov/people/bking/heshop/hemc_papers.html|.
\bibitem{Zholents}    A.A. Zholents,
        {\it The Potential for an Optical Stochastic Cooling After-Burner},
          oral presentation at this workshop. Transparency copies can
          be viewed at~\hfill\break
          \verb|http://pubweb.bnl.gov/people/bking/heshop/hemc_papers.html|.
\bibitem{Lebrun}   Paul Lebrun,
       {\it Comments on Frictional Cooling and the Zero Energy Options
           for Cooling Intense Muon Beams},
          these proceedings. Also available from~\hfill\break 
          \verb|http://pubweb.bnl.gov/people/bking/heshop/hemc_papers.html|.
\bibitem{Nagamine}   Kanetada Nagamine,
        {\it Very Low-Energy Cooling Possibilities Towards Muon Colliders
            and Neutrino Factory}, {\it ibid.}
\bibitem{Carol}   Carol Johnstone,
        {\it Collider Ring Lattices},
          oral presentation at this workshop. Transparency copies can
          be viewed at~\hfill\break
          \verb|http://pubweb.bnl.gov/people/bking/heshop/hemc_papers.html|.
\bibitem{Telnovcomp}   Valery Telnov,
        {\it Some Problems in Plasma Suppression of Beam-Beam Interactions
            at Muon Colliders},
          these proceedings. Also available from~\hfill\break 
          \verb|http://pubweb.bnl.gov/people/bking/heshop/hemc_papers.html|.
\bibitem{Keil}  Eberhard Keil,
        {\it Collective Single-Beam Effects}, {\it ibid.}
\bibitem{Heusch}   Clemens A. Heusch,
         {\it Physics Opportunities with a High-Energy Collider of Same-Sign
               Muons}, {\it ibid.}
\bibitem{epac2000}   B.J. King, parameters and paper in
          preparation for the EPAC 2000 conference.
\end{references}
\end{document}